\documentclass[preprint,superscriptaddress,prb]{revtex4}

\usepackage{graphicx}

\begin{document}

\title{Landau velocity for collective quantum Hall breakdown in bilayer graphene}

\author{W. Yang} \affiliation{Laboratoire Pierre Aigrain, Ecole normale sup\'erieure, PSL University, Sorbonne Universit\'e, Universit\'e Paris Diderot, Sorbonne Paris Cit\'e, CNRS, 24 rue Lhomond, 75005 Paris France}
\author{H. Graef} \affiliation{Laboratoire Pierre Aigrain, Ecole normale sup\'erieure, PSL University, Sorbonne Universit\'e, Universit\'e Paris Diderot, Sorbonne Paris Cit\'e, CNRS, 24 rue Lhomond, 75005 Paris France}
\author{X. Lu}
\affiliation{Beijing National Laboratory for Condensed Matter Physics and Institute of Physics,
Chinese Academy of Sciences, Beijing 100190, China}
\author{G. Zhang}
\affiliation{Beijing National Laboratory for Condensed Matter Physics and Institute of Physics,
Chinese Academy of Sciences, Beijing 100190, China}
\author{T. Taniguchi}
\affiliation{Advanced Materials Laboratory, National Institute for Materials Science, Tsukuba,
Ibaraki 305-0047,  Japan}
\author{K. Watanabe}
\affiliation{Advanced Materials Laboratory, National Institute for Materials Science, Tsukuba,
Ibaraki 305-0047, Japan}
\author{A. Bachtold}
\affiliation{Barcelona Inst Sci and Technol, ICFO Inst Ciencies Foton, Castelldefels 08860,
Barcelona, Spain}
\author{E.H.T. Teo}
\affiliation{Nanyang Technol Univ, Sch Elect and Elect Engn, 50 Nanyang Ave, Singapore 639798, Singapore}
\author{E. Baudin}
\affiliation{Laboratoire Pierre Aigrain, Ecole normale sup\'erieure, PSL University, Sorbonne Universit\'e, Universit\'e Paris Diderot, Sorbonne Paris Cit\'e, CNRS, 24 rue Lhomond, 75005 Paris France}
\author{E. Bocquillon}\affiliation{Laboratoire Pierre Aigrain, Ecole normale sup\'erieure, PSL University, Sorbonne Universit\'e, Universit\'e Paris Diderot, Sorbonne Paris Cit\'e, CNRS, 24 rue Lhomond, 75005 Paris France}
\author{G. F\`eve}\affiliation{Laboratoire Pierre Aigrain, Ecole normale sup\'erieure, PSL University, Sorbonne Universit\'e, Universit\'e Paris Diderot, Sorbonne Paris Cit\'e, CNRS, 24 rue Lhomond, 75005 Paris France}
\author{J-M. Berroir}\affiliation{Laboratoire Pierre Aigrain, Ecole normale sup\'erieure, PSL University, Sorbonne Universit\'e, Universit\'e Paris Diderot, Sorbonne Paris Cit\'e, CNRS, 24 rue Lhomond, 75005 Paris France}
\author{D. Carpentier}
\affiliation{Univ. Lyon, ENS de Lyon, Univ. Claude Bernard, CNRS, Laboratoire de Physique, F-69342,
France}
\author{M. O. Goerbig}
\affiliation{Laboratoire de Physique des Solides, CNRS UMR 8502, Univ. Paris-Sud, Universit\'e
Paris-Saclay, F-91405 Orsay Cedex, France}
\author{B. Pla\c{c}ais} \email{bernard.placais@lpa.ens.fr}
\affiliation{Laboratoire Pierre Aigrain, Ecole normale sup\'erieure, PSL University, Sorbonne Universit\'e, Universit\'e Paris Diderot, Sorbonne Paris Cit\'e, CNRS, 24 rue Lhomond, 75005 Paris France}

\maketitle

\textbf{Breakdown of the quantum Hall effect (QHE) is commonly associated with  an electric field approaching the inter Landau-level (LL) Zener field, ratio of the Landau gap and cyclotron radius. Eluded in semiconducting heterostructures, in spite of extensive investigation, the intrinsic Zener limit is reported here using high-mobility bilayer graphene and high-frequency current noise. We show that collective excitations arising from electron-electron interactions are essential. Beyond a noiseless ballistic QHE regime a large superpoissonian shot noise signals the breakdown via inter-LL scattering. The breakdown is ultimately limited by collective excitations in a regime where phonon and impurity scattering are quenched. The breakdown mechanism can be described by a Landau critical velocity as it bears strong similarities with the roton mechanism of superfluids.}

The Fermi sea of a 2D electronic system is unstable at high magnetic field ($B$-field) toward the formation of discrete Landau levels, giving rise to the quantum Hall effect (QHE) [\onlinecite{Klitzing1980prl}], where the bulk is a Landau insulator. The QHE has led to many developments in physics, including recent ones in metrology [\onlinecite{Tzalenchuk2010nnano,Ribeiro-Palau2015nnano}] and quantum electronics [\onlinecite{Bocquillon2014adp}]. The fate of the QHE at high electric field ($E$-field) remains an open question, as well as the nature of the phase reached when the drift velocity $\mathbf{v}_d=\mathbf{E}\times\mathbf{B}/B^2$ approaches the cyclotron velocity $R_c\omega_c$, where $R_c$ and $\omega_c$ are the cyclotron radius and frequency, resulting in cyclotron orbit breaking. A precursor of the transition is the quantum Hall breakdown reported soon after the discovery of QHE [\onlinecite{Cage1983prl,Streda1984jpc,Bliek1986sst}]. The breakdown field $E_{bd}$ marks the onset of longitudinal resistance and dissipation. Its natural scale is the Zener field, $E_{c}\sim\hbar\omega_c/eR_c$, with $\omega_c=eB/m^*$ and $R_c\sim\sqrt{N}l_B$, where $m^*$ is the effective mass, $N$ the number of occupied LLs  and $l_B=\sqrt{\hbar/eB}$ the magnetic length [\onlinecite{Eaves1986sst}]. The relevance of the Zener mechanism was soon questioned as  $E_{c}$ exceeds experimental $E_{bd}$. Besides, genuine inter-LL tunneling suffers  from a strong momentum mismatch at finite doping,  $\Delta q=2k_F$ where $k_F=\sqrt{2N}/l_B$ is the Fermi momentum [\onlinecite{Dmitriev2012rmp}]. It can be circumvented assuming impurity-assisted or phonon-mediated quasi-elastic inter LL scattering (QUILLS) [\onlinecite{Eaves1986sst,Chaubet1998prb}].

Breakdown was extensively investigated in semiconductors [\onlinecite{Cage1983prl,Bliek1986sst,Kane1988prl,Kaya1999epl,Alexander-Webber2012prb,Li2012prb,Chida2013prb,Panos2014njp}] and  graphene [\onlinecite{Baker2012prb,Guignard2012prb,Alexander-Webber2013prl,Tian2017dm2}], but mainly in  Hall bars. Few experiments used constrictions [\onlinecite{Bliek1986sst,Chida2013prb}], or Corbino geometries [\onlinecite{Chida2014prb,Hata2016jpcm,Laitinen2018scirep,Laitinen2018jltp}], with the purpose of achieving homogeneous current, or $E$-field, distributions. In all cases breakdown $E$-fields are smaller than Zener fields, and critical Hall current densities $J\lesssim 50\;\mathrm{A/m}$ [\onlinecite{Alexander-Webber2013prl}]. The leading explanation thus shifted to a thermal instability, driven by the imbalance between dissipation and phonon relaxation [\onlinecite{Komiyama1986ssc}]; its threshold is material-dependent and lower than $E_{c}$ [\onlinecite{Nachtwei1999physicaE,Komiyama2000prb,Guignard2012prb,Alexander-Webber2013prl}]. According to low-frequency noise measurements, the thermal instability eventually gives rise to electron avalanches [\onlinecite{Chida2013prb,Chida2014prb,Hata2016jpcm}]. On comparing breakdown ($v_{bd}=J_{bd}/ne$) and Zener ($v_c=E_{c}/B$) velocities one finds typically  ratio $v_{bd}/v_c\sim 0.01$ in Hall bars and  $v_{bd}/v_c\lesssim 0.5$ in constrictions. The difference is attributed to the role of electrostatic inhomogeneities in large Hall bars [\onlinecite{Bliek1986sst}], giving rise to peak-effect induced premature breakdowns.

We propose a new scenario where breakdown stems from the spontaneous excitation of large-momentum bulk collective excitations at the magneto-exciton (ME) minimum  [\onlinecite{Kallin1985prb}]. Similar excitations exist in fractional QHE [\onlinecite{Girvin1986prl}] that are longitudinal and called magneto-rotons in reference to the roton minimum in superfluids [\onlinecite{Landau1941pr}]. High momentum excitations take root in short scale correlations due to interactions. They are elusive in spectroscopy, but as suggested by Landau, they can be probed by measuring the critical drift velocity (Landau velocity $v_L$), when their excitation gap, $\omega(q)- v_L q$ in the laboratory frame of reference, vanishes. Recent examples are given by superfluid helium-3 [\onlinecite{Bradley2016nphys}] and quantum gases [\onlinecite{Chomaz2018nphys}]. The case of MEs is generically similar with however some specificities :  the transverse polarization is responsible for an efficient coupling to the  Hall $E$-field; their dispersion doping dependence exhibits a series of $ N$ minima with an oscillator strength vanishing above a cutoff $q\sim k_F$ (see below). With $q_{ME}\sim k_F$ and a rest frequency  $\omega(q_{ME})\sim \omega_c$, magneto-excitons limit the velocity at the Landau velocity $v_L\sim \omega_c /k_F$, which turns out to be identical to the single electron Zener velocity  $v_c=E_{c}/B\sim \omega_c l_B/\sqrt{N}$.
In this work we reach this fundamental limit in a clean bilayer graphene (BLG) micro-sample and reveal the collective nature of breakdown using shot-noise measurements. We report on large  Fano factors $\mathcal{F}\lesssim 20$ and highlight the scaling of shot noise with the Zener field and Landau energy.

Our sample is fabricated from BLG, chosen here as a prototypal  massive ($m^*/m_0\simeq0.03$) 2D electron system with a Landau ladder of inter-LL gap $\hbar\omega_c\lesssim30\;\mathrm{meV}$ for $B\lesssim 8\;\mathrm{T}$. Boron nitride (BN) dielectric provides a high mobility $\mu=3\;\mathrm{m^2 V^{-1}s^{-1}}$ ($T=4\;\mathrm{K}$), a smooth electrostatic environment, and an enhanced thermal stability against the high currents and $E$-fields used in this work [\onlinecite{Yang2017nnano}]. These points are central as they favor homogeneous breakdown over the surface, and high-quality breakdown measurements. A local bottom gate allows tuning the number of occupied LLs in a broad range $N\lesssim10$, while efficiently screening the substrate charge disorder. These statements are supported by several observations. First the electrostatic smoothness is evidenced by the small size of the zero-bias quantization plateaus in Fig.\ref{noise-gallery.fig1}-a and their smearing at low bias $V_{ds}\gtrsim 5\;\mathrm{mV}$. Secondly, the fan chart $\partial G_{ds}(V_{gs},B)/\partial V_g$ in Fig.\ref{noise-gallery.fig1}-b, is characteristic of clean BLG with a low quantizing threshold $B\gtrsim2\;\mathrm{T}$ and a lifting of the fourfold degeneracy of the $N=0$ state. A typical working condition is the point $P$ $\left(B=4\;\mathrm{T}\; ; \; V_g=-3\;\mathrm{V} \right)$ in the middle of the experimental window corresponding to $N\simeq5$ and $n\simeq-2\;10^{12}\;\mathrm{cm^{-2}}$ in Fig.\ref{noise-gallery.fig1}-b. The breakdown is monitored by measuring the transport current and the shot noise. The later is measured at $5\;\mathrm{GHz}$ frequency to overcome low-frequency excess noise which is prominent below $1\;\mathrm{GHz}$ at high current. It is expressed as a noise current $S_I/2e$ in Figs.\ref{noise-gallery.fig1}-(c-j) for an easy comparison with transport currents presented in the same panel.

We report on measurements in a high $E$-field regime unexplored thus far. As shown in Fig.\ref{velocity_noise_scaling.fig2}-a, the QHE regime persists up to large velocities, $v_{bd}\lesssim1.7\;10^5\;\mathrm{m/s}$, approaching the intrinsic limit $v_{bd}\simeq v_c$, which provides a new opportunity to scrutinize the intrinsic mechanisms of breakdown with massive electrons. Remarkably $v_{bd}$ also approaches the phonon saturation velocity, $v_{sat}\simeq2.4\;10^5\;\mathrm{m/s}$ in Ref.[\onlinecite{Yang2017nnano}], which limits the transconductance of graphene field-effect transistors  [\onlinecite{Yang2017nnano,Pallecchi2011apl}]. Specific features of gapless graphene, which can be seen in the $N=0$ state at charge neutrality, are disregarded in this work. They have been investigated recently in single layer graphene [\onlinecite{Laitinen2018jltp}], where breakdown was interpreted in terms of a gyrotropic Zener tunneling effect [\onlinecite{Laitinen2018scirep}].

A representative set of the current and noise measurements in our sample is presented in  Figs.\ref{noise-gallery.fig1},c-j. The bias induced QH breakdown is signaled by a strong uprise of shot noise, widely exceeding the zero $B$-field shot noise, and eventually approaching a full shot noise limit $S_I\approx2eI_{ds}$. The contrast with the ballistic QH regime, where shot noise is suppressed, is striking with a peak noise intensity (at $E\lesssim2E_{bd}$) highlighting a tumultuous breakdown. The onset of noise provides an unambiguous determination of breakdown, because it signals the departure of ballistic transport associated to well-defined Landau levels. The breakdown voltage $V_{bd}(B)=E_{bd} W$, where $W$ is the sample width (see Methods), is indicated by dashed lines in Figs.\ref{noise-gallery.fig1}-(c-j). As seen in the figure, it also corresponds to a deviation from the dissipationless Hall transport regime  $I_{ds}=G_HV_{ds}$ with $G_H=|n| e/B$. \footnote{Note that in a 2-terminal device the electric field rotates from transverse to longitudinal as transport turns from the dissipation-less Hall regime to the dissipative ohmic regime.} In this sample the breakdown currents, $J_{bd}\simeq1200\;\mathrm{A/m}$ (at $B=7\;\mathrm{T}$, $V_g=-6\;\mathrm{V}$), widely exceed previously reported values ($J_{bd}\lesssim50\;\mathrm{A/m}$). The vanishing of noise at  $E\gg E_{bd}$ is different in nature; it signals the ignition of energy relaxation restricting the electronic temperature $k_BT_e\simeq S_I/4G_{ds}$ below a  hot electron limit $k_BT_e\sim eV_{ds}/2$, and indicates the recovery of a metallic behavior.  Conversely, the fact that a full shot noise, or hot electron limit, can be reached at intermediate bias is a direct evidence of the quenching of phonon relaxation mechanisms in our sample. This observation is consistent with the $q=2 k_F$ resonant electron-phonon coupling, reported in phonon-induced resistance oscillations [\onlinecite{Dmitriev2012rmp}], implying a  quenching of conventional phonon relaxation mechanisms at low temperature and high electron density, i.e. in the Bloch-Gruneisen (BG)  regime [\onlinecite{Betz2012prl}]. It also excludes significant contribution from impurity assisted supercollisions [\onlinecite{Betz2013nphys}] in our high mobility sample, intrinsic optical phonon cooling [\onlinecite{Laitinen2014prb}], and demonstrates in-situ the quenching by  quantizing magnetic fields of the highly efficient hyperbolic phonon cooling [\onlinecite{Yang2017nnano}]. Importantly for the interpretation of breakdown, this observation rules out the relevance of the thermal scenario of Refs.[\onlinecite{Nachtwei1999physicaE,Komiyama2000prb,Alexander-Webber2013prl}] or phonon-assisted QUILLS in our experiment which is performed at $4$ Kelvin in the doped regime, i.e. below the BG temperature in conditions where  resonant acoustic phonon scattering is quenched. The fact that supercollisions are not seen at zero field in Ref.[\onlinecite{Yang2017nnano}] is a strong indication that impurity-assisted QUILLS should be very weak. Finally, another salient feature of transport in Figs.\ref{noise-gallery.fig1}-(c-j) is the saturation of the differential conductance at high bias to a value $G_{sat}\simeq 0.6\;\mathrm{mS}$ that is independent of doping and  magnetic field, suggesting that transport becomes metallic. This saturation was analyzed  in terms of Zener-Klein tunneling in Ref.[\onlinecite{Yang2017nnano}].

Signatures of QH breakdown in transport are best captured by the drift velocity $v_d=J_{ds}/ne$ as function of bias expressed in units of the Hall field $E=V_{ds}/W$ in Fig.\ref{velocity_noise_scaling.fig2}-a. It shows a universal Hall mobility, pictured by the bunching of constant $B$-field data along $v_d=E/B=\mu E$ lines. Breakdown is signaled by fanning out of the drift velocity at different densities above $E_{bd}$. The noise determinations of breakdown are more clear cut, they are added in Fig.\ref{velocity_noise_scaling.fig2}-a as black squares, in qualitative agreement with the velocity determination \footnote{Note that bunching quality depends crucially on the accuracy of the drain gating compensation}. Focusing on the median values corresponding to $N=5$ (out of the  experimental window $N = 1\rightarrow10$), the noise data (black squares) can be accurately fitted by the Zener tunneling limit $v_c=E_{c}/B= \omega_cl_B/\sqrt{N}$ (black dashed line) as shown in Fig.\ref{velocity_noise_scaling.fig2}-a. The strong effect of LL quantization on noise is evidenced in Fig.\ref{velocity_noise_scaling.fig2}-b which gathers data of Fig.\ref{noise-gallery.fig1} taken  in the doped regime ($V_g=-3\;\mathrm{V}$), where current noise is expressed in terms of a noise temperature $k_BT_e=S_I/4G_{sat}$. $G_{sat}$ being constant, this representation conveniently interpolates between quantum  shot noise and and thermal noise in increasing bias.  The figure captures salient features of breakdown : the suppression of shot noise below  $E_{bd}$, the proliferation of inter-LL excitations above  $E_{bd}$, with $k_BT_e\sim eV_{ds}/2$, and their relaxation at ultimate bias. As seen in Fig.\ref{velocity_noise_scaling.fig2}-c, the noise data obey themselves a QHE scaling at $E_{bd}$ where $E$-field is scaled to the maximum Zener field, $\tilde{E}=E/E_c$  (obtained taking $N=1$), and the noise temperature to the Landau gap. One retrieves the Zener-like critical field with $\tilde{E}_{bd}\simeq1/\sqrt{5}$ in agreement with the theoretical fit in Fig.\ref{velocity_noise_scaling.fig2}-a. In this plot, we have added  $V_g=0$ data showing that $\tilde{E}_{bd}\rightarrow 1$ at neutrality as expected. The above QHE scaling supports the fundamental physics origin  of the shot noise and highlights two important features :  i) the fact that the Fermi sea is sprayed  over a very large number of LLs ($\gtrsim 100$ in Fig.\ref{velocity_noise_scaling.fig2}-c) under the combined effects of LL quantization and QHE breakdown, and ii) that noise itself scales like the cyclotron energy  with  $S_I(k_F,B,E)\propto\hbar\omega_c\times f(k_F l_B,\tilde{E})$ at ($E\gtrsim E_{bd}$).

 To gain a deeper insight into the breakdown mechanism, we compare in Fig.\ref{superpoissonian.fig3}-a the noise current with the back-scattering current defined as the deviation  from ballistic Hall transport, $I_{bs}=G_HV_{ds}-I_{ds}$. At $E\gtrsim E_{bd}$ the backscattering current identifies with the inter-LL tunneling current which is drained back to the source via a dense array of counter-propagating edge states. Both $S_I$ and $I_{bs}$ vanish in the QHE regime $E< E_{bd}$ and grow exponentially at $E\sim E_{bd}$.  As $S_I(V_g)\propto I_{bs}(V_g)$, we can define a $E$-field and doping independent backscattering Fano factor  $\mathcal{F}(B)=S_I/ 2eI_{bs}$. At $B=3\;\mathrm{T}$ the large value $\mathcal{F}\simeq 7.5$ (deduced from the ratio of the noise current and backscattering current in the  Fig.\ref{superpoissonian.fig3}-a) points to a collective mechanism and rules out the single electron Zener interpretation of breakdown. The inset of Fig.\ref{superpoissonian.fig3}-a shows the linear dependence $\mathcal{F}(B)\propto B$ which supports the above QHE scaling of noise. It introduces a constant energy scale, $\varepsilon_c=\hbar\omega_c/\mathcal{F}\simeq1.4\;\mathrm{meV}$, the origin of which will be discussed below.

From a theoretical point of view, the critical $E$-field for the onset of breakdown can be understood as due to a proliferation of magneto-excitons. These collective excitations arise as zeros in the dielectric function $\epsilon[E,\omega({\bf q})]=0$, which, before breakdown, satisfies Galilean invariance $\epsilon[E,\omega({\bf q})]=\epsilon[E=0, \omega_{E=0}({\bf q}) - {\bf v}_D\cdot {\bf q}]=0$, where $\omega_{E=0}({\bf q})$ is the dispersion of the collective excitation at rest, i.e. in the absence of an electric field. We concentrate on magneto-excitons [\onlinecite{Kallin1985prb}] that slightly modulate the cyclotron frequency as a function of the wave vector, providing $N$ minima in the range $1/\sqrt{2N}l_B\leq q \leq 2\sqrt{2N+1}/l_B=2k_F$ of maximal spectral weight [\onlinecite{Roldan2010,Goerbig2011rmp}]. The position of the last ME minimum, which determines the critical drift velocity, therefore scales as $q_{ME}\sim \sqrt{N}/l_B$, and its energy is roughly given by $\hbar \omega_{ME}\simeq \hbar \omega_c (1 + \beta r_s/\sqrt{N})$. Here, the doping-dependant Seitz radius $r_s=1/a_B^* k_F\simeq 0.5...1$ is given by the effective Bohr radius $a_B^*= 0.5 {\rm \AA} \times (m_e/m^*)\epsilon$, with $\epsilon=3.2$ for BN. The offset $\beta \omega_c r_s/\sqrt{N}$ ($\beta$ is an unimportant numerical prefactor) is therefore a fraction of the cyclotron frequency that we neglect in our orders-of-magnitude argument. In order to evaluate the critical $E$-field for proliferation of magneto-excitons, we thus determine the Landau critical velocity $v_L q_{ME} \sim \omega_c$, which yields the above scaling law $v_L\sim \omega_c l_B/\sqrt{N}$.
Fig.\ref{superpoissonian.fig3}-b is a sketch of the low energy ME spectrum $\omega_{ME}(q)-v_d q$ in increasing  drift velocities $v_d=0$, $0.5$ and $1\;10^5\;\mathrm{m/s}$ for our typical experimental parameters $|n|=2\;10^{12}\;\mathrm{cm^{-2}}$, $B=4\;\mathrm{T}$ ($N=5$). Line thickness sketches ME's oscillator strength which, according to Refs.[\onlinecite{Roldan2010}], vanishes for $q\gtrsim k_F$. The ME instability occurs at a Landau critical velocity $v_L\sim \omega_c/k_F\simeq1\;10^5\;\mathrm{m/s}$, at the intersection with the zero energy line, in agreement with the experimental ($4\;\mathrm{T}$) value in Fig.\ref{velocity_noise_scaling.fig2}-a where $v_{bd}=1.1\;10^5\;\mathrm{m/s}$.

The above $E$-field-induced magneto-exciton instability signals the proliferation of inter-LL collective excitations. Hence it explains the typical breakdown $E$-field and the large Fano factor of the noise. However a more quantitative determination of the number $N_{bunch}=\mathcal{F}\lesssim20$ of excitation bunches is beyond the scope of the present argument. This would require a comprehensive theory of the polarisability at high $E$-field and of the contribution to electric transport of the collective excitations. Moreover, experimental feedback  would require a knowledge of the bias-dependent Hall angle which is not accessible in 2-terminal RF measurements. Still we can estimate the electrostatic back-action from the inter-edge charging energy $\varepsilon_c\sim e^2/(\epsilon W)\simeq 2\;\mathrm{meV}$, assuming an instantaneous bulk to edge counter drift of electron/hole partners after an elementary ME creation. In this simple picture,  the local  $E$-field is reduced below the threshold under magneto-exciton production whenever the Coulomb potential $N_{bunch}\varepsilon_c$ compensates the Landau gap, i.e.  $N_{bunch}=\mathcal{F}\sim\hbar\omega_c/\varepsilon_c$ in order of magnitude accordance with experiment. This relaxation oscillation mechanism bears analogies with the Frank-Condon mechanism in nanoelectronics where giant Fano factors are considered under phonon-triggered tunneling events  [\onlinecite{Koch2005prl}].

Transport and noise measurements provide hints that QH Landau insulator turns to a metallic state at very high electric fields ($E\gtrsim2E_{bd}$) with a differential conductance  $G_{ds}\rightarrow G_{sat}\simeq0.6\;\mathrm{mS}$ in Fig.\ref{noise-gallery.fig1}-(c-j) and shot noise $S_I(B,E)\rightarrow S_I(0,E)$ in Fig.\ref{velocity_noise_scaling.fig2}-c. This trend is a natural consequence of the field-induced hybridization of LLs. Remarkably, it is accompanied by the ignition of hyperbolic substrate-phonon cooling, an efficient mechanism  recently demonstrated in the same sample [\onlinecite{Yang2017nnano}]. The suppression of phonon relaxation in quantizing fields, also reported in magneto-optic experiments [\onlinecite{Wendler2017prl}], and its restoring upon LL merging, are an experimental illustration of the importance of the plane wave nature of electrons in electron-phonon coupling. Modeling this coupling would require accessing to the infra-red optical conductivity spectrum which remains beyond the scope of the present work.

In conclusion, the old problem of quantum Hall breakdown has been revisited in the light of high-frequency shot-noise measurements performed in high-mobility bilayer graphene. A new mechanism has been proposed, which involves an electric-field driven bulk magneto-exciton instability. It explains  the observed transport and noise phenomenology as well as their scaling with  Zener field and Landau energy. The mechanism makes a bridge between quantum Hall effect and superfluid breakdowns by introducing a Landau critical drift velocity to the Landau insulator state.  This new approach will stimulate further theoretical and experimental works, aiming at modeling the $E$-field-induced QHE to metal transition or checking its universality, e.g. in single-layer graphene where Lorenzian invariance substitutes Galilean invariance, in ultra-clean semiconducting heretostructures, or in fractional quantum Hall effect. Finally, the proposed mechanism for bulk QHE breakdown may inspire an edge-state variant that would be highly relevant for quantum Hall metrology.

\section{Methods}

\begin{figure}[hh]
\centerline{\includegraphics[width=15cm]{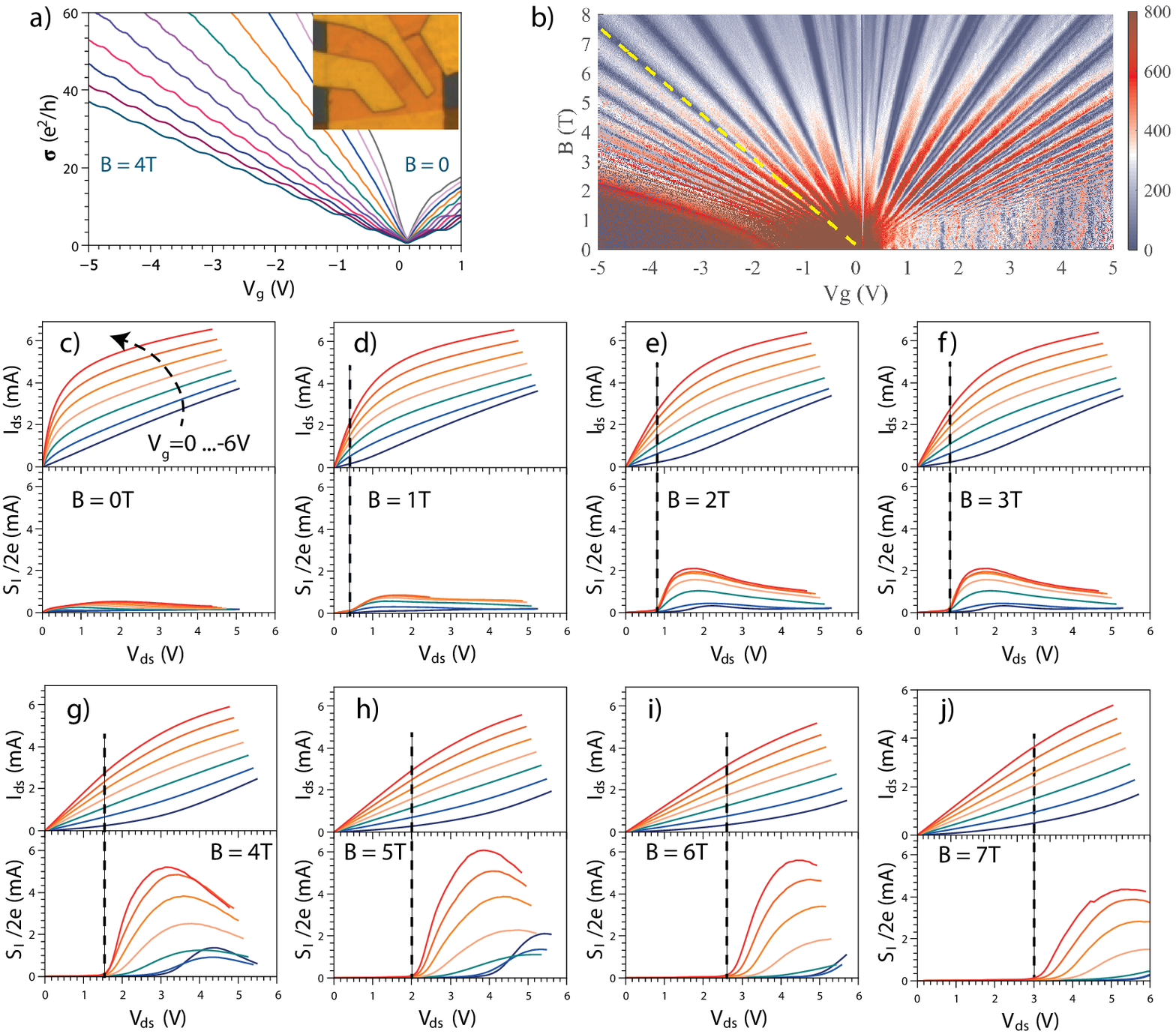}}
\newpage
\caption{Magnetotransport and high electric-field shot noise in bilayer graphene at $T=4\;\mathrm{K}$.
a) Low $E$-field conductance quantization steps (at $V_{ds}=3\;\mathrm{mV}$ in units of $4e^2/h$) of the 2-terminal bottom gated bilayer graphene on $23\;\mathrm{nm}$-thick boron nitride (BN) device of dimensions $L\times W=4\times3\;\mathrm{\mu m}$ (optical image in the inset). The tiny width of the quantum Hall plateaus warrants the absence of disorder-induced localized bulk states.
b) Fan chart of the zero-field differential conductance $\partial G_{ds}/\partial V_{g}$  ($\mu S/V$) showing a series of Landau levels ($N=-10\rightarrow10$, the yellow dashed line signals the median $N=5$ state) and the lifting of their 4-fold degeneracy of the $N=0$ state.
c-j) High-bias transport current $I_{ds}(V_{ds})$ and  shot-noise current $S_I/2e(V_{ds})$ (measured at $5\;\mathrm{GHz}$) in the QH regime at negative gate voltage, $V_g=0\rightarrow-6\;\mathrm{V}$, corresponding to hole doping, $|n|=0$--$4.3\;10^{12}\;\mathrm{cm^{-2}}$. At quantizing fields ($B\gtrsim2\;\mathrm{T}$) shot noise is suppressed at low bias where $G_{ds}=|n|e/B$, rises up abruptly at a breakdown voltage (dashed black lines), peaks to a large value eventually reaching the full shot-noise limit $S_I\approx2e I_{ds}$ ($B=5\;\mathrm{T}$), and finally vanishes as conductance saturates to a field-independent value $G_{sat}\simeq0.6\;\mathrm{mS}$. }
 \label{noise-gallery.fig1}
\end{figure}

\begin{figure}[hh]
\centerline{\includegraphics[width=15cm]{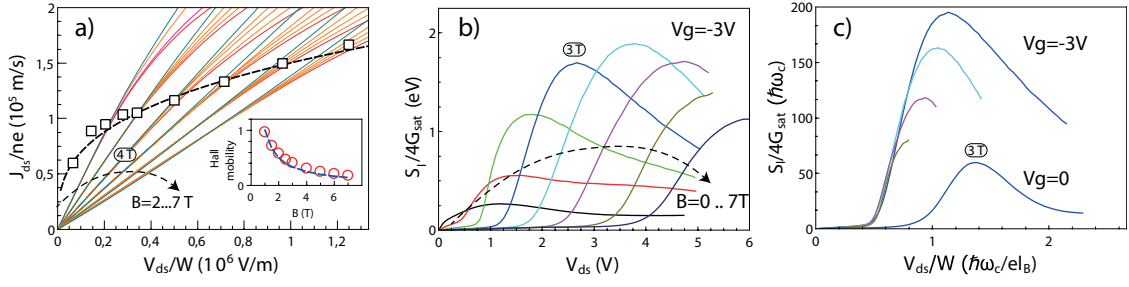}}
\caption{Scaling behavior of current and noise in the QHE regime.
a) Bias dependence of the drift velocity $v_d=J_{ds}/|n|e$ deduced from current data in Fig.\ref{noise-gallery.fig1}. Taken at fixed magnetic field, the data of different carrier densities bunch together in the QH regime and stray above the breakdown field. Different bunches correspond to $B=2\rightarrow7\;\mathrm{T}$ magnetic fields. The slopes of the bunches, pictured as a Hall mobility in the inset,  obey the Hall mobility  $\mu=1/B$  (blue dashed line in $m^2/Vs$). Black squares correspond to a typical breakdown voltage deduced from the onset of shot noise (dashed lines in Fig.\ref{noise-gallery.fig1}-(c-j). They can be fitted to the Zener tunneling limit $v_c=\left(E_{c}\hbar e/Nm^{*2}\right)^{1/3}$ (black dashed line) by taking a constant landau level $N=5$ corresponding to the median experimental value (dashed yellow line in Fig.\ref{noise-gallery.fig1}-b).
b) Bias dependence of shot noise in increasing magnetic fields $B=0\rightarrow7\;\mathrm{T}$ at a fixed carrier density $|n|=2.15\;10^{12}\;\mathrm{cm^{-2}}$. Data are reproduced from Fig.\ref{noise-gallery.fig1}-(e-j).  Ballistic transport is characterized by a suppression of shot noise below a critical voltage ranging from $V_c\simeq0.6\;\mathrm{V}$ (at $B=2\;\mathrm{T}$) to  $V_c\simeq4\;\mathrm{V}$ (at $B=7\;\mathrm{T}$). Current noise is scaled by the constant saturation conductance $G_{sat}=0.6\;\mathrm{mS}$ to be expressed as a noise temperature.
c)  the above noise data obey a quantum Hall scaling at breakdown where Hall voltage and noise temperature are  scaled to the Zener field $E_c=\hbar\omega_c/el_B$ and cyclotron energy $\hbar\omega_c$ respectively. The breakdown occurs at a reduced scaled field $\tilde{E}_{bd}\simeq 1/\sqrt{N}\simeq 0.45$ according to the  median landau level number $N=5$. The neutrality data are shown for comparison with a reduced scaled field $\tilde{E}_{bd}\sim 1$. }
\label{velocity_noise_scaling.fig2}
\end{figure}

\begin{figure}[ht]
\centerline{\includegraphics[width=15cm]{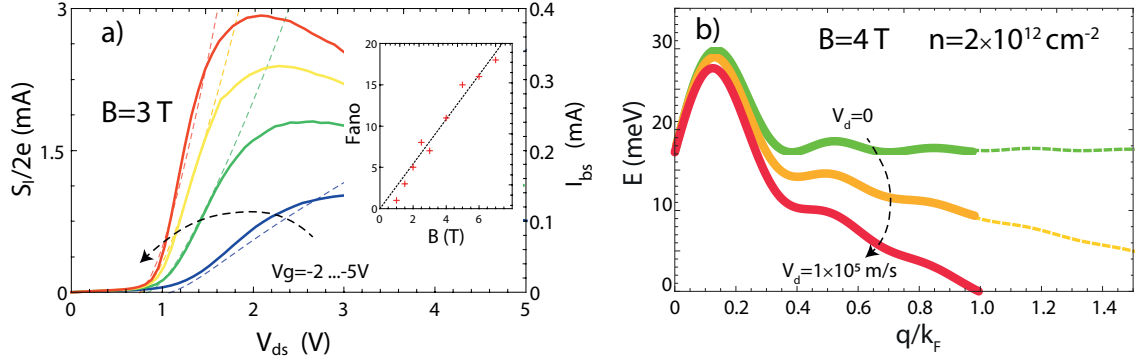}}
\caption{Superpoissonian inter Landau level tunneling and a sketch of the magneto-exciton instability scenario of QHE breakdown.
a) Shot noise (solid lines, left axis) and backscattering current $I_{bs}=G_HV_{ds}-I_{ds}$ (dashed lines, right axis) at $B=3\;\mathrm{T}$ for $V_g=-2\rightarrow-5\;\mathrm{V}$ (blue, green, orange, red) obey a similar bias dependence at breakdown yielding a doping independent ratio, the  Fano factor $\mathcal{F}=S_I/2eI_{bs}\simeq7.5$. Inset : the magnetic field dependence of the Fano factor obeys a linear law $\mathcal{F}(B)=I\simeq2.7B\approx \hbar\omega_c/\varepsilon_c$ with $\varepsilon_c=1.4\;\mathrm{meV}$.
b) Magneto-exciton (ME) spectrum, $E=\hbar\omega_c\left[1+(Ak_F/q)\;J_1[2Nq/k_F]^2\right]-\hbar v_dq$, at rest ($v_d=0$), for a subcritical ($v_d=0.5\;10^5\;\mathrm{m/s}$) and the critical ($v_d=1\;10^5\;\mathrm{m/s}$ drift velocities.  Parameters are  $A=0.7$, $B=4\;\mathrm{T}$ and $|n|=2\;10^{12}\;\mathrm{cm^{12}}$ ($N=5$), corresponding to the $V_g=-3\;\mathrm{V}$ median working point in Fig.\ref{noise-gallery.fig1}-b. The line thickness illustrates the oscillator strength which, according  to Ref.[\onlinecite{Roldan2010}], vanishes at $q\sim k_F$. The Landau critical velocity $v_L=1\;10^5\;\mathrm{m/s}$ is estimated by  the vanishing of the ME gap at $q/k_F\sim1$. It agrees with the noise determination of breakdown, $v_{bd}=1.1\;10^5\;\mathrm{m/s}$  ($B=4\;\mathrm{T}$  square in Fig.\ref{velocity_noise_scaling.fig2}-a).}
\label{superpoissonian.fig3}
\end{figure}

 The 2-terminal sample, of dimensions  $L\times W=4\times3\;\mathrm{\mu m}$, is an as-exfoliated BLG flake stacked on a $23\;\mathrm{nm}$-thick BN crystal deposited on a metallic bottom gate  (see optical picture in Fig.\ref{noise-gallery.fig1}-a inset).  The sample is equipped with low-resistance ($R_c\simeq 120\;\mathrm{\Omega.\mu m}$) Pd/Au contacts, and embedded in a coplanar wave guide for $0$--$10\;\mathrm{GHz}$ cryogenic ($4\;\mathrm{K}$) noise measurement. Shot noise is measured in the $4.5$--$5.5\;\mathrm{GHz}$ band to overcome the low frequency resistance noise up to the highest bias currents [\onlinecite{Yang2017nnano}]. The high-frequency band  is needed to overcome the colored low frequency noise that obscures shot noise up to $\sim1\;\mathrm{GHz}$ at the maximum bias current. Carrier density is compensated for drain gating as explained in [Ref.\onlinecite{Yang2017nnano}]. The dissipation-less Hall regime is characterized by a 2-terminal conductance $G_{ds}$ matching the Hall conductance $G_H=|n|e/B$.  QHE plateaus are used to calibrate the gate capacitance $C_g\simeq1.15\;\mathrm{mF/m^2}$.  The $B=0$  transport and noise properties have been characterized in a previous report (Ref.[\onlinecite{Yang2017nnano}]) where a new cooling mechanism could be revealed and experimental techniques were detailed.

\begin{acknowledgments}
The research leading to these results have received partial funding from the the European Union ``Horizon 2020'' research and innovation programme under grant agreement No. 785219 ``Graphene Core'',  from the ANR-14-CE08-018-05 "GoBN". G.Z. acknowledges the financial supports from the National Basic Research Program of China (973 Program) under the grant No. 2013CB934500, the National Science Foundation of China (NSFC) under the grant No. 61325021.
\end{acknowledgments}

\textbf{Author contributions} WY and BP conceived the experiment and developed the models.
WY conducted the measurements. WY, XL, TT, KW, GZ participated to sample fabrication. WY, GF,
JMB, EB, AB, ET and BP participated to the data analysis. MOG, DC and HG worked out the theoretical modeling.  WY, MOG and BP wrote the manuscript with contributions from the coauthors.

\textbf{Additional information} Competing financial interests: The authors declare no competing
financial interests.

\newpage

\end{document}